\documentclass[prd,showpacs]{revtex4}

\usepackage{bm}
\usepackage{amsmath}
\usepackage{natbib}

\def\ba{{\bm a}}
\def\bb{{\bm b}}

\def\bl{{\bm l}}

\def\bN{{\bm N}}

\def\bx{{\bm x}}

\def\bla{{\bm \lambda}}
\def\bbe{{\bm \beta}}

\def\bs{\mbox{\boldmath $s$}}

\def\lb{\label}

\def\be{\begin{equation}}
\def\ee{\end{equation}}
\def\bea{\begin{eqnarray}}
\def\eea{\end{eqnarray}}

\begin{document}

\title{Angular distances in metric theories}

\author{Pierre Teyssandier}
\email{Pierre.Teyssandier@obspm.fr}
\affiliation{D\'epartement Syst\`emes de R\'ef\'erence Temps et Espace,
CNRS/UMR 8630, \\
Observatoire de Paris, 61 avenue de l'Observatoire, F-75014 Paris, France}
 
\author{Christophe Le Poncin-Lafitte}
\email{christophe.le_poncin-lafitte@tu-dresden.de}
\affiliation{Lohrmann Observatory, Dresden Technical University,\\ Mommsenstr. 13, D-01062 Dresden, Germany}

\date{\today}

\begin{abstract}
The general expression of the angular distance between two point sources as measured by an arbitrary observer is given. The modelling presented here is rigorous, covariant and valid in any space-time. The sources of light may be located at a finite distance from the observer. The aberration and the gravitational deflection of light are treated in a unified way. Assuming the gravitational field to be weak, an explicit expansion of the angular separation within the post-post-Minkowskian approximation is carried out. The angular separation within the post-Newtonian approximation truncated at the order $1/c^3$ is straightforwardly derived.
\end{abstract}


\maketitle

\section{Introduction}

An accuracy of the order of the microarcsecond ($\mu$as) is expected in future space astrometry missions, like the Global Astrometric Interferometer for Astrophysics (GAIA) which is planned to be launched at the latest in 2012 \cite{GAIA:2000,Perryman:2001,Bienayme:Turon:2002}, or the Space Interferometry Mission (SIM) \cite{sim}. Several models of positional observations compatible with such a level of accuracy have been proposed within the post-Newtonian approximation of general relativity (see, e.g., \cite{Klioner:1992} and \cite{defelice:2006}). An even more general model was formulated by Klioner \cite{Klioner:2003a} in the framework of the parametrized post-Newtonian (PPN) formalism \cite{Will:1993}, with parameters $\beta$ and $\gamma$. Such models are sufficient for the currently envisaged astrometry space missions. However, quite ambitious laser interferometric missions like LATOR \cite{lator} and ASTROD \cite{astrod} are now proposed with the aim of testing the post-linear approximation of metric theories near the Sun. The main purpose of the present work is to outline a rigorous and general modelling of the angular distance between two light sources as measured by an observer having a given motion and then to specialize to the different approximations which will be needed in the foreseeable future.

All the perturbation expansions performed in this paper are deduced from the rigorously covariant formulae given in Sec. II. It must be emphasized that the aberration and the gravitational deflection of light are not separately treated. Moreover, the sources of light (supposed to be pointlike) may be located at finite distances from the observer. As a consequence, all the formulae given here take into account the motion of the observer and are suitable both for remote stars or quasars and for objects moving in the solar system.   
 
In Sec. II we derive the general expression of the angular separation between two point light sources as measured by a given observer. We use in this section a coordinate-free approach totally avoiding the index notation owing to the simplicity of the calculations. In Sec. III we show that the angular separation can be explicitly calculated when the reception time transfer function as defined in \cite{leponcin:2004} is determined provided that the motion of the observer is known. In Sec. IV, we get the general expansion allowing to determine the angular separation within the post-post-Minkowskian approximation. In Sec. V, we carry out the calculations within the post-Newtonian approximation. Section VI delivers some concluding remarks.

Throughout this work, $c$ is the speed of light in a vacuum and $G$ is the Newtonian gravitational constant. The Lorentzian metric of space-time $V_4$ is denoted by $g$. We adopt the signature $(+ - - -)$. As long as explicit coordinates are not needed, four dimensional vectors are denoted by usual letters : $u, v, l, l', U, \beta_{u/v}, \beta_{v/u}, ...$. We put $u.v = g(u, v) = g_{\alpha\beta}u^{\alpha} v^{\beta}$ and $u^2 = u.u$.

When an explicit introduction of local coordinates is required, we suppose that space-time is covered by some global coordinate system $x^{\alpha} = (x^0, \bx)$, with $x^0 = ct$ and $\bx = (x^i)$. Moreover, we assume that the curves of equations $x^i =$ const are timelike in the neighbourhood of the observer. This condition means that $g_{00}> 0$ in the vicinity of the observer. We employ the usual three-dimensional vector notation $\ba$ in order to denote either the ordered set $(a^1, a^2 , a^3)$, or the orderer set $(a_1, a_2 , a_3)$. Since there cannot exist any confusion with the above-mentioned notation of the scalar product of two 4-vectors, given  $\ba = (a^1, a^2 , a^3)$, for instance, we denote by $\ba . \bb$ the quantity $a^i b^i$ if $\bb = (b^1, b^2 , b^3)$ and $a^i b_i$ if $\bb = (b_1, b_2 , b_3)$), the Einstein convention of summation on repeated indices being used in each case. The quantity $\vert \ba \vert$ denotes the ordinary Euclidean norm of $\ba$ : $\vert \ba \vert = (\delta_{ij}a^i a^j)^{1/2}$ if $\ba = (a^1, a^2 , a^3)$, and 
$\vert \ba \vert = (\delta^{ij}a_i a_j)^{1/2}$ if $\ba = (a_1, a_2 , a_3)$. The indices between parentheses characterize the order of a term in a perturbative expansion. Theses indices are set up or down, depending on the convenience.

When there exists a risk of ambiguity, we use the subscript $o$ for quantities related to the observer : for example, $x_{o}$ is the point-event where the observation is located. We use the subscript $e$ for quantities related to the emission of a light ray : $t_{e}$ is the coordinate time of emission and $\bx_{e}$ denotes the position of the emitter.

\section{Angular distance as measured  by a given observer}

Let us consider a light ray $\Gamma$ received at point $x_{o}=(ct_{o}, {\bm x}_{o})$ and let us recall how is defined the direction of this ray as measured by an observer ${\cal O}(u)$ moving at $x_{o}$ with a unit 4-velocity $u$. The three-space relative to the observer ${\cal O}(u)$ at point $x_{o}$ is the subspace $\Pi_{x_{o}}^{(3)}(u)$ of tangent vectors orthogonal to $u$. An arbitrary vector $V$ at $x_{o}$ admits one and only one decomposition of the form
\be \lb{pr0}
V = V_{\parallel_{u}} + V_{\bot_{u}} \, , 
\ee
where $V_{\parallel_{u}}$ is colinear to the unit vector $u$ and $V_{\bot_{u}}$ is a vector of the three-space $\Pi_{x_{o}}^{(3)}(u)$. Since $V_{\bot_{u}}$ and $u$ are orthogonal, one has 
\be \lb{pr1}
V_{\parallel_{u}} = (u . V) u
\ee
and 
\be \lb{pr2}
V_{\bot_{u}} = V - (u . V) u \, .
\ee
The vector $V_{\bot_{u}}$ is called the (orthogonal) projection of $V$ onto the three-space relative to the observer ${\cal O}(u)$. Its magnitude $\mid\!V_{\bot_{u}}\!\mid = \sqrt{- V_{\bot_{u}} . V_{\bot_{u}}}$ is given by
\be \lb{le1}
\mid\!V_{\bot_{u}}\!\mid  = \sqrt{(u . V)^2 - V^2} \, .
\ee

The direction of vector $V$ as seen by the observer ${\cal O}(u)$ is the direction of the unit spacelike vector $V_{\bot_{u}}^{\ast}$ defined as
\be \lb{di1}
V_{\bot_{u}}^{\ast} = \frac{V_{\bot_{u}}}{\mid\!V_{\bot_{u}}\!\mid} = \frac{V - (u . V) u }{\sqrt{(u . V)^2 - V^2}} \, .
\ee

Consider now a light ray $\Gamma$ received at $x_{o}$ and denote by $l$ a vector tangent to $\Gamma$ at $x_{o}$. In this work, we always assume that a vector tangent to a light ray is a null, future directed vector, so that
\be \lb{nul}
l^2 = 0 \, , \qquad \quad u.l >0 \, .
\ee

The direction of the ray $\Gamma$ as measured by the observer ${\cal O}(u)$ is the direction of the vector $l_{\bot_{u}}^{\ast}$. By using Eq. (\ref{di1}), and then taking into account Eqs. (\ref{nul}), it is easy to see that   
\be \lb{di2}
l_{\bot_{u}}^{\ast} = \frac{l}{u . l} - u \, .
\ee

Let $\Gamma'$ be another light ray received at $x_{o}$. If $l'$ denotes a vector tangent to $\Gamma'$ at $x_{o}$, the direction of $\Gamma'$ as observed by ${\cal O}(u)$ is given by Eq. (\ref{di2}) in which $l'$ is substituted for $l$. As a consequence, the angular separation between $\Gamma$ and $\Gamma'$ as measured by ${\cal O}(u)$ may be defined as the angle $\phi_u$ between the two vectors $l_{\bot_{u}}^{\ast}$ and $l_{\bot_{u}}'^{\ast}$ belonging to the same subspace $\Pi_{x_{o}}^{(3)}(u)$ (see, e.g., \cite{Soffel:1988,brumberg:1991}). The angle $\phi_u$ may be characterized without ambiguity by relations as follow
\be \lb{an1}
\cos \phi_{u} = - \, l_{\bot_{u}}^{\ast} . \, l_{\bot_{u}}'^{\ast} \, , \qquad 0 \leq \phi_{u} \leq \pi \, .
\ee

By using Eq. (\ref{di2}) and then taking into account that $u$ is a unit timelike vector, it is easily deduced from (\ref{an1}) that
\be \lb{an2}
\cos \phi_u = 1 - \frac{l . \, l'}{(u . l)(u . l') } \, ,
\ee

As a consequence, the angle $\phi_{u}$ is determined by 
\be \lb{an3}
\sin^2 \frac{\phi_u}{2} = \frac{1}{2} \frac{l . l'}{(u . l)(u . l')} \, , \qquad 0 \leq \phi_{u} \leq \pi \, .
\ee
Since $l$ and $l'$ are supposed to be null vectors, we have
\be \lb{ll}
l.l' = - \frac{1}{2}(l'-l)^2 \, .
\ee
By substituting for $l.l'$ from Eq. (\ref{ll}) into Eq. (\ref{an3}), we get the formula
\be \lb{an4}
\sin^2 \frac{\phi_u}{2} = - \frac{1}{4} \frac{(l' - l)^2}{(u . l)(u . l')} \, .
\ee

Let us now compare $\phi_{u}$ with the angular distance $\phi_v$ between $\Gamma$ and $\Gamma'$ as measured by an observer moving with a unit 4-velocity $v$. It follows from (\ref{an3}) that
\be \lb{ab1}
\sin^2 \frac{\phi_u}{2} = \frac{(v . l)(v . l')}{(u . l)(u . l')} \, \sin^2 \frac{\phi_v}{2}\, .
\ee

The factor ahead of $\sin^2 \frac{\phi_v}{2}$ is the aberration factor due to the motion of the observer ${\cal O}(u)$ relative to the observer ${\cal O}(v)$. Let us determine this factor in a more explicit form.

Denote by $\tau$ the proper time of the observer ${\cal O}(u)$. The spatial displacement of ${\cal O}(u)$ in the 3-space $\Pi^{3}_{x_o}(v)$ relative to ${\cal O}(v)$ between instants $\tau$ and $\tau + d\tau$ is the vector
$$
dx_{\bot_v} = c \left[u - (u . v)v  \right]d\tau \, .
$$
This displacement corresponds to a timelike displacement in a frame comoving with ${\cal O}(v)$ given by 
$$
dx_{\parallel_v} = c(u . v) \, v d\tau \, .
$$
As a consequence, the velocity vector $w_{u/v}$ of the observer ${\cal O}(u)$ relative to the observer ${\cal O}(v)$ may be defined as 
\be \lb{ve1}
w_{u/v} = c \beta_{u/v}\, ,
\ee
where $\beta_{u/v}$ is the vector given by (see also \cite{bolos})
\be \lb{veb}
\beta_{u/v} = \frac{dx_{\bot_v}}{\mid\!dx_{\parallel_v}\!\mid} =  \frac{u}{u . v} - v  \, .
\ee 

The magnitude of the spacelike vector $\beta_{u/v}$ has an expression as follows
\be \lb{sp1}
\mid\!\beta_{u/v}\!\mid = \sqrt{- \beta_{u/v}^2} = \sqrt{1 - \frac{1}{(u . v)^2}} \, .
\ee

This last relation is equivalent to 
\be \lb{sp2}
\frac{1}{(u . v)^2} = 1 - \mid\!\beta_{u/v}\!\mid^2 \, .
\ee

So, $u.v$ is the $\gamma$ factor corresponding to the velocity $\vert w_{u/v} \vert$.
 
Conversely, the velocity vector of the observer ${\cal O}(v)$ relative to the observer ${\cal O}(u)$ is given by
\be \lb{ve2}
w_{v/u} = c \beta_{v/u} \, , \qquad \beta_{v/u} =  \frac{v}{u . v} - u  \, .
\ee

It results from Eq. (\ref{sp1}) that $\mid\!\beta_{u/v})\!\mid$ and $\mid\!\beta_{v/u}\!\mid$ have the same magnitude :
\be \lb{sp}
\mid\!\beta_{v/u}\!\mid = \mid\!\beta_{u/v}\!\mid \, ,
\ee
a property which is well known in special relativity. Let us note, however, that $\beta_{u/v}$ and $\beta_{v/u}$ cannot be opposite vectors since they do not belong to the same subspace, except in the case where $u = v$ at point $x_o$.

We are now in a position to write Eq. (\ref{ab1}) in a more explicit form involving $\beta_{u/v}$ and the unit spacelike vectors $l_{\bot_{v}}^{\ast}$ and $l_{\bot_{v}}'^{\ast}$ which characterize the directions of the rays $\Gamma$ and $\Gamma'$, respectively, in the three-space relative to the observer ${\cal O}(v)$. Indeed, $l$ and $l'$ may be written as 
\be \lb{l}
l = (v . l) (v + l_{\bot_{v}}^{\ast}) \, , \qquad  l' = (v . l') (v + l_{\bot_{v}}'^{\ast}) \, 
\ee
and Eq. (\ref{veb}) yields
\be \lb{u}
u = (u.v) (v + \beta_{u/v}) \, .
\ee
Since $l_{\bot_{v}}^{\ast}$, $l_{\bot_{v}}'^{\ast}$ and $\beta_{u/v}$ are orthogonal to $v$, Eqs. (\ref{l}) and (\ref{u}) imply 
\be \lb{ull}
u.l = (u.v) (v.l)(1 + \beta_{u/v} . l_{\bot_{v}}^{\ast})\, , \qquad u . l' = (u.v) (v.l')(1 + \beta_{u/v} . l_{\bot_{v}}'^{\ast})\, .
\ee
Substituting for $u.l$ and $u.l'$ from their respective expressions given by Eqs. (\ref{ull}) into Eq. (\ref{ab1}) and taking Eq. (\ref{sp2}) into account yields the fundamental relation
\be \lb{ab2}
\sin^2 \frac{\phi_u}{2} = \frac{1 - \mid\!\beta_{u/v}\!\mid^2}{\left(1 + \beta_{u/v} . l_{\bot_{v}}^{\ast} \right)\left(1 + \beta_{u/v} . l_{\bot_{v}}'^{\ast} \right)} \, \sin^2 \frac{\phi_v}{2} \, ,
\ee
which gives $\phi_{u}$ as a function of $\phi_{v}$ and of vectors which are all in the three-space $\Pi_{x_{o}}^{(3)}(v)$ relative to the observer ${\cal O}_{v}$.

Since the null vectors $l$ and $l'$ are assumed to be future directed, the directions in which the observer ${\cal O}(v)$ is seeing the sources of the light rays $\Gamma$ and $\Gamma'$ are characterized by the unit vectors $N_{v}$ and $N_{v}'$ defined as
\be \lb{N1}
N_{v} = - \, l_{\bot_{v}}^{\ast} \, , \qquad N_{v}' = - \, l_{\bot_{v}}'^{\ast} \, .
\ee

Of course, $\phi_v$ is the angle between $N_{v}$ and $N_{v}'$.

Denote by $\psi_v$ (resp. $\psi_{v}'$) the angle between $\beta_{u/v}$ and $N_v$ (resp. $\beta_{u/v}$ and $N_{v}'$). One has  
\bea 
& &\mid\!\beta_{u/v}\!\mid  \cos \psi_v = - N_v . \beta_{u/v} = \beta_{u/v} . l_{\bot_{v}}^{\ast} \, , \qquad  0\leq \psi_{v} \leq \pi \, , \lb{ps1} \\
& &\mid\!\beta_{u/v}\!\mid \cos \psi_{v}' = - N_{v}' . \beta_{u/v} = \beta_{u/v} . l_{\bot_{v}}'^{\ast} \, , \qquad  0\leq \psi_{v}' \leq \pi \, . \lb{ps2}
\eea
Then Eq. (\ref{ab2}) may be rewritten as 
\be \lb{ab3}
\sin^2 \frac{\phi_u}{2} = \frac{1 - \mid\!\beta_{u/v}\!\mid^2}{(1+\mid\!\beta_{u/v}\!\mid  \cos \psi_v)(1+\mid\!\beta_{u/v}\!\mid  \cos \psi_{v}')}
\, \sin^2 \frac{\phi_v}{2} \, .
\ee

Conversely, put
\be \lb{N2}
N_{u} = - \, l_{\bot_{u}}^{\ast} \, , \qquad N_{u}' = - \, l_{\bot_{u}}'^{\ast} \, ,
\ee
and define the angles $\psi_{u}$ and $\psi_{u}'$ by
\bea 
& &\mid\!\beta_{v/u}\!\mid  \cos \psi_{u} = - N_u . \beta_{v/u} = \beta_{v/u} . l_{\bot_{u}}^{\ast} \, , \qquad  0\leq \psi_{u} \leq \pi \, ,\lb{ps3} \\
& &\mid\!\beta_{v/u}\!\mid \cos \psi_{u}' = - N_{u}' . \beta_{v/u} = \beta_{v/u} . l_{\bot_{u}}'^{\ast} \, , \qquad  0\leq \psi_{u}' \leq \pi \, . \lb{ps4}
\eea
We may write
\be \lb{ab4}
\sin^2 \frac{\phi_v}{2} = \frac{1 - \mid\!\beta_{v/u}\!\mid^2}{(1+\mid\!\beta_{v/u}\!\mid  \cos \psi_u)(1+\mid\!\beta_{v/u}\!\mid  \cos \psi_{u}')}
\, \sin^2 \frac{\phi_u}{2} \, .
\ee

It is worthy of note that comparing Eqs. (\ref{ab3}) and (\ref{ab4}) yields an identity as follows :
\be \lb{id}
\frac{(1+\mid\!\beta_{v/u}\!\mid  \cos \psi_u)(1+\mid\!\beta_{v/u}\!\mid  \cos \psi_{u}')}{1 - \mid\!\beta_{v/u}\!\mid^2}
= \frac{1 - \mid\!\beta_{u/v}\!\mid^2}{(1+\mid\!\beta_{u/v}\!\mid  \cos \psi_v)(1+\mid\!\beta_{u/v}\!\mid  \cos \psi_{v}')} \, .
\ee

\section{Angular separation expressed in a given coordinate system}

Up to now, we have not specified any coordinate system. Henceforth, we shall assume that the coordinate system $x^{\alpha} = (x^0, x^i)$ is chosen so that $\partial_{0} \equiv \partial/\partial x^0$ is a timelike vector in the neighbourhood of the point of observation. This condition means that $g_{00} >0$ in the vicinity of $x_o = (x_o^0, x_o^i)$. On this assumption, we can consider that the curve defined by $x^i = x_o^i = $ const is the worldline of an observer ${\cal O}(U)$ at rest relative to the reference system defined by the coordinate system $x^{\alpha}$ and passing through $x_{o}$. At $x_o$, this observer is moving with a unit 4-velocity $U$ given by
\be \lb{U}
U = \frac{1}{\sqrt{g_{00}}} \frac{\partial}{\partial x^0} \qquad \Longleftrightarrow \qquad U^{\alpha} = \frac{\delta^{\alpha}_{0}}{\sqrt{g_{00}}} \, .
\ee

A null vector cannot be orthogonal to a timelike vector. As a consequence, the null vectors $l$ and $l'$ are such that
\be \lb{ll1}
l_0 = (l . \partial_{0}) \neq 0 \, , \qquad l'_0 = (l' . \partial_{0}) \neq 0 \, .
\ee
So the above formulae remain valid under the substitutions
\be \lb{sub}
l \longrightarrow \widehat{l}\, , \qquad l' \longrightarrow \widehat{l}'\, ,
\ee
where $\widehat{l}$ and $\widehat{l}'$ are the null vectors defined by
\be \lb{ll2}
\widehat{l} = \frac{l}{(l.\partial_{0})} \, , \qquad
\widehat{l}' = \frac{l'}{(l'.\partial_{0})} \, .
\ee

The introduction of vectors $\widehat{l}$ and $\widehat{l}'$ will be justified below.

Using these definitions, the angular separation $\phi_U$ between $\Gamma$ and $\Gamma'$ as measured by the observer ${\cal O}(U)$ is given by a relation as follows
\be \lb{an5} 
\sin^2 \frac{\phi_U}{2} = - \frac{1}{4}\frac{(\widehat{l}' - \widehat{l})^2}{(U. \widehat{l}) (U.\widehat{l}')} \, ,
\ee 
and the relation between $\phi_u$ and $\phi_U$ may be written as
\be \lb{ab5}
\sin^2 \frac{\phi_u}{2} = \frac{(U . \widehat{l})(U . \widehat{l}')}{(u . \widehat{l})(u . \widehat{l}')} \, \sin^2 \frac{\phi_U}{2}\, ,
\ee

Noting that the covariant components of $\widehat{l}$ and $\widehat{l}'$ are given by
\be \lb{ll3}
\widehat{l}_0 = 1\, , \,  \widehat{l}_i = \frac{l_i}{l_0}\, , \qquad  \widehat{l}_0' = 1\, , \,  \widehat{l}_i' = \frac{l_i'}{l_0'} \, ,
\ee
and then, taking into account Eq. (\ref{U}), it is easy to see that Eq. (\ref{an5}) reads
\be \lb{an6}
\sin^2 \frac{\phi_U}{2} = - \frac{1}{4} \left[g_{00} g^{ij}(\widehat{l}_{i}' - \widehat{l}_{i})(\widehat{l}_{j}' - \widehat{l}_{j})\right]_{x_o} \, .
\ee

Then, applying the general formula (\ref{ab3}) with $v= U$ yields
\be \lb{ab6}
\sin^2 \frac{\phi_u}{2} = - \frac{1}{4} \frac{1 - \mid\!\beta_{u/U}\!\mid^2}{(1+\mid\!\beta_{u/U}\!\mid  \cos \psi_U)(1+\mid\!\beta_{u/U}\!\mid  \cos \psi_{U}')}
\, \left[g_{00} g^{ij}(\widehat{l}_{i}' - \widehat{l}_{i})(\widehat{l}_{j}' - \widehat{l}_{j})\right]_{x_o} \, .
\ee

In many practical problems, however, it will be indispensable to express all the factors involved in $\sin^2 \frac{\phi_u}{2}$ in terms of coordinate-dependent quantities. Of course, the contravariant components of the vector $\beta_{u/U}$ could be explicited by replacing $v$ by $U$ in Eq. (\ref{veb}), and then using Eq. (\ref{U}). Nevertheless, using Eqs. (\ref{U}) and (\ref{ab5})-(\ref{an6}) yields straightforwardly
\be \lb{an7}
\sin^2 \frac{\phi_u}{2} = - \frac{1}{4}\left[\frac{1}{(u^0)^2} \,  \frac{g^{ij}(\widehat{l}_{i}' - \widehat{l}_{i})(\widehat{l}_{j}' - \widehat{l}_{j})}{(1+ \beta^{m}\, \widehat{l}_{m})(1+\beta^{r}\,\widehat{l}_{r}')}\right]_{x_{o}} \, ,
\ee
where quantities $\beta^{i}$ are the components of the coordinate-velocity of ${\cal O}(u)$ divided by $c$ :
\be \lb{hlb}
\beta^{i} = \frac{dx^i}{dx^0} = \frac{dx^i}{c dt}  \, . 
\ee

Since $u$ is a unit timelike vector, one has 
\be \lb{u0}
\frac{1}{(u^0)^2} = g_{00} + 2 g_{0i} \beta^{i} + g_{ij}\beta^{i}\beta^{j} \, .
\ee

Finally, substituting for $1/(u^0)^2$ from Eq. (\ref{u0}) into Eq. (\ref{an7}) yields the fundamental formula 
\be \lb{an8}
\sin^2 \frac{\phi_u}{2} = - \frac{1}{4} \left[  
\frac{\left( g_{00} + 2 g_{0k} \beta^{k} + g_{kl}\beta^{k}\beta^{l} \right) g^{ij}(\hat{l}'_{i} - \hat{l}_{i})(\hat{l}'_{j} - \hat{l}_{j})}
{(1+ \beta^{m}\, \hat{l}_{m})(1+\beta^{r}\,\hat{l}'_{r})} 
\right]_{x_{o}} \, .
\ee

This formula gives the expression $\phi_u$ as a function of the metric, the coordinate velocity of the observer ${\cal O}(u)$ and the six quantities $\widehat{l}_i$ and $\widehat{l}_i'$ at point $x_{o}$. In order to determine the quantities $\widehat{l}_i$, we may proceed as follows. Suppose that the light ray $\Gamma$ received at point $(ct_{o}, {\bm x}_{o})$ is emitted at point $(ct_e, {\bm x}_{e})$. The travel time $t_{o} - t_{e}$ of this ray may be considered as a function of ${\bm x}_{e}$, $t_{o}$ and ${\bm x}_{o}$. So we can put  
\be \lb{tmt}
t_{o}-t_{e} = {\mathcal T}_{r}({\bm x}_{e}, t_{o}, {\bm x}_{o})\,,
\end{equation}
where ${\mathcal T}_{r}$ may be called the {\em reception time transfer function}. Then, as it is proved in \cite{leponcin:2004}, we may write 
\begin{equation} \label{cov}
\left(\hat{l}_{i}\right)_{x_o} = \left(\frac{l_{i}}{l_{0}}\right)_{x_o} =  
- c \, \frac{\partial  {\cal T}_{r}({\bm x}_{e}, t_{o}, {\bm x}_{o})}{\partial x^{i}_{o}}
\left[1 - \frac{\partial  {\cal T}_{r}({\bm x}_{e}, t_{o}, {\bm x}_{o})} {\partial t_{o}}\right]^{-1}.
\ee 

As a consequence, once the reception time transfer function is explicitly known in a given space-time, it is possible to obtain the angular separation as a function of the coordinates of the emission and reception points, and of the coordinate speed of the observer.

Henceforth, we focus our attention on the case of a weak gravitational field. So we suppose that space-time is covered by some global quasi-Galilean coordinate system $x^{\alpha} = (x^0, \bx)$ in which the metric may be written as
\be \lb{gmn}
g_{\mu \nu} = \eta_{\mu \nu} + h_{\mu \nu} \, , \qquad \mbox{$\eta_{\mu \nu}$ = diag $(1, -1, -1, -1)$} \, ,
\ee
where the gravitational perturbations $h_{\mu\nu}$ meet the conditions $\vert h_{\mu \nu} \vert <<1$. According to this assumption, the time transfer function ${\mathcal T}_r$ may be written in the form
\be \lb{Tr}
{\mathcal T}_{r}({\bm x}_{e}, t_{o}, {\bm x}_{o}) = \frac{1}{c} \vert \bx_e - \bx_o \vert + \Theta_{r}({\bm x}_{e}, t_{o}, {\bm x}_{o})\, .
\ee

Substituting for ${\mathcal T}_{r}$ from Eq. (\ref{Tr}) into (\ref{cov}) and then putting
\be \lb{N}
N^{i} = \frac{x_{e}^{i} - x_{o}^{i}}{\mid\! \bx_{e} - \bx_{o}\!\mid}\, , \qquad N'^{i} = \frac{x'^{i}_{e} - x_{o}^{i}}{\mid\! \bx'_{e} - \bx_{o}\!\mid}\, ,
\ee
it is easily seen that 
\be \lb{hl2}
\left(\hat{l}_{i}\right)_{x_{o}} = N^{i} + \lambda_i (\bx_{e}, t_o, \bx_{o}) \, , \quad \left(\hat{l}'_{i}\right)_{x_{o}} = N'^{i} + \lambda_i (\bx'_{e}, t_o, \bx_{o}) \, ,
\ee
where the functions $\lambda_i (\bx, t_o, \bx_{o})$ are defined as
\be \lb{lamb}
\lambda_i (\bx, t_o, \bx_{o}) = - \left[1 - \frac{\partial \Theta_{r}({\bm x}, t_{o}, {\bm x}_{o})}{\partial t_{o}} \right]^{-1}
\left[c\, \frac{\partial  \Theta_{r}({\bm x}, t_{o}, {\bm x}_{o})}{\partial x_{o}^{i}} - \frac{\partial  \Theta_{r}({\bm x}, t_{o}, {\bm x}_{o})}{\partial t_{o}}\, \frac{x^{i} - x_{o}^{i}}{\mid\! \bx - \bx_{o}\!\mid} \right] \, .
\ee 

The functions $\lambda_i (\bx, t_o, \bx_{o})$ characterize the fact that a light ray emitted at point $\bx_{e}$ is not arriving at $(ct_{o}, \bx_{o}$ in the direction defined by the direction cosines $(x^{i} - x_{o}^{i})/\mid\! \bx_{e} - \bx_{o}\!\mid$. So we call them {\em the (gravitational) deflection functions}. 

For the sake of brevity, we introduce the vector notations
\bea 
& &\hat{\bl} = \{ \hat{l}_{i} \}\, ,\qquad \hat{\bl'} = \{ \hat{l}_{i}' \}\, , \lb{ll4} \\
& &\bN = \{N^{i} \} = \frac{\bx_{e} - \bx_{o}}{\vert \bx_{e} - \bx_{o}\vert} \, , \qquad  \bN' = \{N'^{i} \} = \frac{\bx_{e}' - \bx_{o}}{\vert \bx_{e}' - \bx_{o}\vert} \, , \lb{NN} \\
& &\bla = \bla(\bx_{e}, t_o, \bx_{o}) = \{\lambda_{i}(\bx_{e}, t_o, \bx_{o}) \} \, ,\qquad \bla' = \bla(\bx_{e}', t_o, \bx_{o}) = \{\lambda_{i}(\bx_{e}', t_o, \bx_{o}) \}\, . \lb{ll5}
\eea

Thus Eqs. (\ref{hl2}) may be written in the form
\be \lb{hbl}
\hat{\bl} = \bN + \bla \, , \qquad \hat{\bl}' = \bN' + \bla' \, .
\ee

It is clear that $\bN$ is the zeroth-order, unit coordinate vector from the observer to the source. Noting that $ \vert \bN' - \bN \vert = 
\sqrt{2(1 - \bN . \bN')}$, we define a ``unit vector" $\bs$ characterizing the angular separation as 
\be \lb{si}
\bs = \{s^{i}\} = \frac{\bN' - \bN}{\sqrt{2(1 - \bN . \bN')}} \, .
\ee

Finally, we put
\be \lb{vec}
\bbe = \{ \beta^{i} \} \, , \qquad \beta^2 = \delta_{ij} \beta^{i}\beta^{j}\, .   
\ee

Substituting for $\hat{l}_{i}$ and $\hat{l}'_{i}$ from Eqs. (\ref{hl2}) into Eq. (\ref{an5}), and then using the above-mentioned notations, we get 
\bea \lb{an9}
\sin^2 \frac{\phi_u}{2} &=& \frac{1}{2}\, \frac{1 - \beta^{2} + h_{00} + 2 h_{0k}\beta^{k} + h_{kl}\beta^k \beta^l}{[1 + \bbe . (\bN + \bla)][1 + \bbe . (\bN' + \bla')]}  \nonumber \\
& & \nonumber \\
& &\times \left\{ \left(1 - \bN . \bN'\right) \left(1 - k^{ij}s^i s^j \right)  + \sqrt{2(1- \bN . \bN')} \, \left[(\bla' - \bla). \bs -
 k^{ij} (\lambda'_{i} - \lambda_{i}) s^{j} \right] \right. \nonumber \\
& &\left. \qquad \qquad \qquad \qquad + \frac{1}{2} (\bla' - \bla)^2 - \frac{1}{2} k^{ij} (\lambda'_{i} - \lambda_{i})(\lambda'_{j} - \lambda_{j})        \right\} \, ,
\eea
where the quantities $k^{\mu\nu}$ are defined by 
\be \lb{kmn}
g^{\mu \nu} = \eta^{\mu \nu} + k^{\mu \nu} \, .
\ee

Let us emphasize that the formula (\ref{an9}) is rigorous.

\section{Angular distance within the post-post-Minkowskian approximation}

Let us assume now that the metric perturbations $h_{\mu\nu}$ may be represented at any point $x$ by a series in ascending powers of the Newtonian gravitational constant $G$ 
\be \lb{hmn}
h_{\mu \nu}(x, G) = \sum_{1}^{\infty} G^n g_{\mu \nu}^{(n)}(x) \, .
\ee

We shall put 
\be \lb{h}
h_{\mu \nu}^{(n)} = G^n \, g_{\mu \nu}^{(n)}(x) \, .
\ee

More generally, given any function $f(x, G)$ which admits an expansion as follows
\be \lb{fun}
f(x, G) = \sum_{n=0}^{\infty} G^n u^{(n)}(x) \, ,
\ee
we shall put for the sake of brevity
\be \lb{fn}
f^{(n)}(x, G) = G^n u^{(n)}(x) \, ,
\ee
so that $f$ may be written as
\be \lb{ffn}
f(x, G) = \sum_{n=0}^{\infty} f^{(n)}\, .
\ee

For the sake of convenience, we shall sometimes use the notation $f_{(n)}$ instead of $f^{(n)}$.

In what follows, we restrict our attention to the post-post-Minkowskian approximation, which means that we neglect all terms involving $h_{\mu \nu}^{(3)}, h_{\mu \nu}^{(4)}, ...$ As a consequence $k^{ij}$ may be replaced by 
\be \lb{kij}
k^{ij} = k_{(1)}^{ij} + k_{(2)}^{ij} + O(G^3) \, ,
\ee
where
\bea 
k_{(1)}^{ij} &=& -\, h_{ij}^{(1)} \, , \lb{kij1} \\
k_{(2)}^{ij} &=& -\, h_{ij}^{(2)} + \eta^{\alpha \beta} h_{i \alpha}^{(1)}h_{j \beta}^{(1)} \, . \lb{kij2}
\eea

According to Eq. (\ref{hmn}), we assume that the vector deflection functions $\bla$ and $\bla'$ admit expansions as follows 
\be \lb{la}
\bla = \bla^{(1)} + \bla^{(2)} + O(G^3) \, , \qquad \bla' = \bla'^{(1)} + \bla'^{(2)} + O(G^3) \, .
\ee

Furthermore, we point out that in all practical cases, the speed of the observer is much smaller than the celerity of light in a vacuum. So we henceforth assume that 
\be \lb{bet}
0 \leq \bbe^2 < 1 \, ,
\ee 
which implies that a condition as follows
\be \lb{abe}
(1 + \bbe . \bN)(1 + \bbe . \bN') \neq 0 
\ee
is satisfied.

Using the previous expansions, a long but straightforward calculation shows that the angle $\phi$ is determined by an expansion as follows
\bea \lb{a10}
\sin^2 \frac{\phi_u}{2} &=& \frac{1}{2}\, \frac{1 - \beta^{2}}{(1 + \bbe . \bN)(1 + \bbe . \bN')} \left\{ \left(1 - \bN . \bN'\right)\left[ 1 + 
{\cal A}^{(1)} + {\cal A}^{(2)}\right] \qquad \qquad \qquad \qquad \qquad \right. \nonumber \\
& & \left. \qquad \qquad \qquad \qquad \qquad \quad  + \sqrt{2\left(1 - \bN . \bN'\right)} \left[ {\cal B}^{(1)} + {\cal B}^{(2)}\right] + \frac{1}{2}
\left( \bla'^{(1)} - \bla^{(1)} \right)^2 + O(G^3)\right\}\, ,
\eea
where
\bea \lb{A1}
{\mathcal A}^{(1)} &=& \frac{h_{00}^{(1)} + 2 h_{0i}^{(1)}\beta^{i}  + h_{ij}^{(1)}\beta^i \beta^j}{1 - \beta^{2}} + h_{ij}^{(1)}s^i s^j 
- \frac{\bbe . \bla^{(1)}}{1 + \bbe . \bN} - \frac{\bbe . \bla'^{(1)}}{1 + \bbe . \bN'} \, ,  \\
& & \nonumber \\
{\mathcal B}^{(1)} &=& \left(\bla'^{(1)} - \bla^{(1)}\right).\, \bs \lb{B1} \, ,
\eea
and
\bea \lb{A2}
{\mathcal A}^{(2)} &=& \frac{h_{00}^{(2)} + 2 h_{0i}^{(2)}\beta^{i} + h_{ij}^{(2)}\beta^i \beta^j}{1 - \beta^{2}} - 
\frac{h_{00}^{(1)} + 2 h_{0i}^{(1)}\beta^{i} + h_{ij}^{(1)}\beta^i \beta^j}{1 - \beta^{2}} \left( \frac{\bbe . \bla^{(1)}}{1 + \bbe . \bN} + \frac{\bbe . \bla'^{(1)}}{1 + \bbe . \bN'} \right) \nonumber \\
& &+ \left[ h_{ij}^{(2)} + \left( \frac{h_{00}^{(1)} + 2 h_{0k}^{(1)}\beta^{k} + h_{kl}^{(1)}\beta^k \beta^l}{1 - \beta^{2}} - 
\frac{\bbe . \bla^{(1)}}{1 + \bbe . \bN} - \frac{\bbe . \bla'^{(1)}}{1 + \bbe . \bN'} \right)h_{ij}^{(1)} - \eta^{\gamma \delta}h_{i\gamma}^{(1)}h_{j\delta}^{(1)}  \right]s^i s^j \nonumber \\
& &- \frac{\bbe . \bla^{(2)}}{1 + \bbe . \bN} - \frac{\bbe . \bla'^{(2)}}{1 + \bbe . \bN'} + 
\left(\frac{\bbe . \bla^{(1)}}{1 + \bbe . \bN}\right)^2 + \frac{\bbe . \bla^{(1)}}{1 + \bbe . \bN} \cdot \frac{\bbe . \bla'^{(1)}}{1 + \bbe . \bN'} + \left(\frac{\bbe . \bla'^{(1)}}{1 + \bbe . \bN'}\right)^2 \, , \\
& & \nonumber \\
& & \nonumber \\
{\mathcal B}^{(2)} &=& \left[\bla'^{(2)} - \bla^{(2)} + \left( \frac{h_{00}^{(1)} + 2 h_{0k}^{(1)}\beta^{k} + h_{kl}^{(1)}\beta^k \beta^l}{1 - \beta^{2}} - 
\frac{\bbe . \bla^{(1)}}{1 + \bbe . \bN} - \frac{\bbe . \bla'^{(1)}}{1 + \bbe . \bN'}\right)\left(\bla'^{(1)} - \bla^{(1)}\right) \right].\, \bs \nonumber \\
& & \qquad \qquad \qquad \qquad \qquad \qquad + h_{ij}^{(1)}\left(\bla'^{(1)}_{i} - \bla_{i}^{(1)} \right)s^j \, . \lb{B2}
\eea

Denote by $\phi_u^{(0)}$ the angle defined by 
\be \lb{an0}
\sin^2 \frac{\phi_{u}^{(0)}}{2} = \frac{1}{2}\, \frac{1 - \beta^{2}}{(1 + \bbe . \bN)(1 + \bbe . \bN')} \left(1 - \bN . \bN'\right) \, , 
\quad 0 \leq \phi_u^{(0)} \leq \pi \, .
\ee

It is clear that $\phi_{u}^{(0)}$ is the angle which would be measured by the observer ${\cal O}(u)$ in the absence of any gravitational field. As a consequence it may be assumed that $\phi$ admits an expansion as follows  
\be \lb{phi}
\phi_{u} = \phi_{u}^{(0)} + \phi_{u}^{(1)} + \phi_{u}^{(2)} + O(G^3) \, .
\ee 

Using Taylor's theorem, it may be seen that
\be \lb{tay}
\sin^2 \frac{\phi_u}{2} = \sin^2 \frac{\phi_{u}^{(0)}}{2} + \frac{1}{2}\, \phi_{u}^{(1)}\sin \phi_{u}^{(0)} + \frac{1}{2}\, \phi_{u}^{(2)}\sin \phi_{u}^{(0)}
+ \frac{1}{4}\left(\phi_{u}^{(1)}\right)^2\cos \phi_{u}^{(0)} + O(G^3) \, .
\ee

Now compare Eq. (\ref{tay}) with Eq. (\ref{an7}), and then take Eqs. (\ref{A1})-(\ref{B2}) into account. We find for the first-order perturbation term  
\be \lb{ph1}
\phi_{u}^{(1)} =  {\mathcal A^{(1)}} \tan\frac{\phi_{u}^{(0)}}{2} +  \frac{\sqrt{1 - \beta^{2}}}{\sqrt{(1 + \bbe . \bN)(1 + \bbe . \bN')}} \left(\cos \frac{\phi_{u}^{(0)}}{2}\right)^{-1} {\mathcal B^{(1)}} \, , 
\ee
or, more explicitly
\bea \lb{ph1b}
\phi_{u}^{(1)} &=& \tan\frac{\phi_{u}^{(0)}}{2} \left[ \frac{h_{00}^{(1)} + 2 h_{0i}^{(1)}\beta^{i} + h_{ij}^{(1)}\beta^i \beta^j}{1 - \beta^{2}} + h_{ij}^{(1)}s^i s^j 
- \frac{\bbe . \bla^{(1)}}{1 + \bbe . \bN} - \frac{\bbe . \bla'^{(1)}}{1 + \bbe . \bN'} \right] \nonumber \\
& &+ \frac{\sqrt{1 - \beta^{2}}}{\sqrt{(1 + \bbe . \bN)(1 + \bbe . \bN')}}\left(\cos \frac{\phi_{u}^{(0)}}{2}\right)^{-1} 
\left(\bla'^{(1)} - \bla^{(1)}\right).\, \bs \, .
\eea

For the second-order term, we obtain
\bea \lb{phi2}
\phi_{u}^{(2)} &=& \tan\frac{\phi_{u}^{(0)}}{2} \, {\mathcal C^{(2)}} +  \frac{\sqrt{1 - \beta^{2}}}{\sqrt{(1 + \bbe . \bN)(1 + \bbe . \bN')}}
\left(\cos \frac{\phi_{u}^{(0)}}{2}\right)^{-1} {\mathcal D^{(2)}}  \nonumber \\
& & + \frac{1}{2} \frac{1 - \beta^{2}}{(1 + \bbe . \bN)(1 + \bbe . \bN')}\, \frac{1}{\sin \phi_{u}^{(0)}}
\left\{ \left( \bla'^{(1)} - \bla^{(1)} \right)^2 - \left(1 - \tan^{2}\frac{\phi_{u}^{(0)}}{2}\right) 
\left[ \left(\bla'^{(1)} - \bla^{(1)}\right).\, \bs\right]^2 \right\}\, ,
\eea
where
\bea \lb{C2}
{\mathcal C^{(2)}}&=& {\mathcal A^{(2)}} - \frac{1}{4} \left(1 - \tan^{2}\frac{\phi_{u}^{(0)}}{2}\right)\left({\mathcal A^{(1)}}\right)^2 \nonumber \\
&=&\frac{h_{00}^{(2)} + 2 h_{0i}^{(2)}\beta^{i} + h_{ij}^{(2)}\beta^i \beta^j}{1 - \beta^{2}} - \frac{1}{4}\, \left(1 - \tan^{2}\frac{\phi_{u}^{(0)}}{2}\right)
\left[ \left(\frac{h_{00}^{(1)} + 2 h_{0i}^{(1)}\beta^{i} + h_{ij}^{(1)}\beta^i \beta^j}{1 - \beta^{2}}\right)^2 + \left( h^{(1)}_{ij}s^i s^j\right)^2 \right]
\nonumber \\
& &- \frac{1}{2}\left(1 + \tan^{2}\frac{\phi_{u}^{(0)}}{2}\right)\frac{h_{00}^{(1)} + 2 h_{0i}^{(1)}\beta^{i} + h_{ij}^{(1)}\beta^i \beta^j}{1 - \beta^{2}}
 \left( \frac{\bbe . \bla^{(1)}}{1 + \bbe . \bN} + \frac{\bbe . \bla'^{(1)}}{1 + \bbe . \bN'} \right) \nonumber \\
& &+ \left[ h_{ij}^{(2)} +  \frac{1}{2}\left(1 + \tan^{2}\frac{\phi_{u}^{(0)}}{2}\right) \left( \frac{h_{00}^{(1)} + 2 h_{0k}^{(1)}\beta^{k} + h_{kl}^{(1)}\beta^k \beta^l}{1 - \beta^{2}} - 
\frac{\bbe . \bla^{(1)}}{1 + \bbe . \bN} - \frac{\bbe . \bla'^{(1)}}{1 + \bbe . \bN'} \right)h_{ij}^{(1)} - \eta^{\gamma \delta}h_{i\gamma}^{(1)}h_{j\delta}^{(1)}  \right] s^i s^j \nonumber \\
& &- \frac{\bbe . \bla^{(2)}}{1 + \bbe . \bN} - \frac{\bbe . \bla'^{(2)}}{1 + \bbe . \bN'} + \frac{1}{4}\left(3 + \tan^{2}\frac{\phi_{u}^{(0)}}{2}\right)
\left[ \left(\frac{\bbe . \bla^{(1)}}{1 + \bbe . \bN}\right)^2 + \left(\frac{\bbe . \bla'^{(1)}}{1 + \bbe . \bN'}\right)^2 \right] \nonumber \\
& & \qquad \qquad \qquad \qquad \qquad \qquad + \frac{1}{2}\left(1 + \tan^{2}\frac{\phi_{u}^{(0)}}{2}\right)
\frac{\bbe . \bla^{(1)}}{1 + \bbe . \bN} \cdot \frac{\bbe . \bla'^{(1)}}{1 + \bbe . \bN'}  \, , 
\eea
and 
\bea \lb{D2}
{\mathcal D^{(2)}} &=& {\mathcal B^{(2)}} - \frac{1}{2} \left(1 - \tan^{2}\frac{\phi_{u}^{(0)}}{2}\right){\mathcal A^{(1)}} {\mathcal B^{(1)}} \nonumber \\
&=&\left(\bla'^{(2)} - \bla^{(2)}\right) . \, \bs + \frac{1}{2}\left[ \left(1 + \tan^{2}\frac{\phi_{u}^{(0)}}{2}\right)\right. \left( \frac{h_{00}^{(1)} + 2 h_{0k}^{(1)}\beta^{k} + h_{kl}^{(1)}\beta^k \beta^l}{1 - \beta^{2}} - 
\frac{\bbe . \bla^{(1)}}{1 + \bbe . \bN} - \frac{\bbe . \bla'^{(1)}}{1 + \bbe . \bN'}\right)  \nonumber \\
& &\left. \qquad \qquad \qquad \qquad \qquad - \left(1 - \tan^{2}\frac{\phi_{u}^{(0)}}{2}\right)
h^{(1)}_{ij} s^i s^j \right]\left[\left(\bla'^{(1)} - \bla^{(1)}\right) .\, \bs \right] + h_{ij}^{(1)}\left(\bla'^{(1)}_{i} - \bla_{i}^{(1)} \right)s^j \, .
\eea

{\em Post-Minkowskian approximation}. The previous formulae drastically simplify if we neglect all the contributions of second order with respect to $G$. Let us introduce the unperturbed angular distance $\phi_U^{(0)}$ which would be measured by the observer ${\cal O}(U)$ at rest relative to the chosen coordinate system in the absence of gravitational field :
\be \lb{anr}
\sin^2 \frac{\phi_U^{(0)}}{2} = \frac{1}{2} (1 - \bN . \bN') \, , \qquad 0 \leq \phi_{U}^{(0)} \leq \pi \, .
\ee

We obtain within the linearized, weak field approximation :
\bea \lb{a11}
\sin^2 \frac{\phi_u}{2} &=& \frac{1}{2}\, \frac{1 - \beta^{2}}{(1 + \bbe . \bN)
(1 + \bbe . \bN')} \left\{ (1 - \bN . \bN')\left[1 + \frac{h_{00} + 2 h_{0i}\beta^{i} + h_{ij}\beta^i \beta^j}{1 - \beta^{2}} + h_{ij}s^i s^j \right. \right.
\nonumber \\
& &\left. \left. - \, \frac{\bbe . \bla}{1 + \bbe . \bN} - \, \frac{\bbe . \bla'}{1 + \bbe . \bN'}  \right]  + \sqrt{2(1- \bN . \bN')} \,\, (\bla' - \bla) . \, \bs  + O(G^2)\right\} \, .
\eea

Assuming $\phi_U^{(0)}\neq 0$ and $\vert\beta \vert <<1$ allow to develop the square root of the r.h.s. of (\ref{a11}) as follows :
\bea \lb{a12}
\sin \frac{\phi_{u}}{2} &=& \frac{\sqrt{1 - \beta^{2}}}{\sqrt{\left(1 + \bbe . \bN \right)\left(1 + \bbe . \bN' \right)}}
\left\{\sin \frac{\phi_{U}^{(0)}}{2}\left[ 1 + \frac{1}{2}\frac{h_{00} + 2 h_{0i}\beta^{i} + h_{ij}\beta^i \beta^j}{1 - \beta^{2}} + \frac{1}{2}
h_{ij} s^i s^j \right. \right. \nonumber \\
& &\left. \left. - \frac{1}{2} \frac{\bbe . \bla}{1 + \bbe . \bN} - \frac{1}{2}\frac{\bbe . \bla'}{1 + \bbe . \bN'}\right] + \frac{1}{2} \left(\bla' - \bla \right).\,\bs + O(G^2) \right\} \, .
\eea

\section{Angular distance within the post-Newtonian approximation}

Let us now suppose that the weak gravitational field considered in the previous section is generated by a system of $N$ self-gravitating bodies within the slow-motion, post-Newtonian approximation. By choosing a convenient quasi Cartesian coordinate system (e.g., a standard post-Newtonian gauge), the metric may be written in the form
\bea
&&h_{00} = - \frac{2w}{c^2}+O(c^{-4})\, , \lb{pn1} \\
&&h_{0i} =  O(c^{-3})\, ,\lb{pn2} \\
&&h_{ij} = 2\gamma\frac{w}{c^2}\eta_{ij}+ O(c^{-4})\, , \lb{pn3}
\eea
where $w$ is given by 
\be \lb{w}
w(x^0, \bx) = G \sum_{a=1}^{N}\int_{{\cal D}_b} \frac{\rho_{a}(x^0, \bx')}{\vert\bx - \bx' \vert} d^3\bx' + O(c^{-2}) \, ,
\ee
$\rho_{a}$ being the rest mass density of the body $a$ and ${\cal D}_{a}$ the domain occupied by the body $a$ at instant $x^0$.
 
Within this approximation, we may neglect all the terms which are of the order of $1/c^4$ and we have to retain only the terms of order $1/c^2$ in the gravitational deflection functions $\lambda_i$ and $\lambda_{i}'$. Substituting for $h_{\mu\nu}$ from Eqs. (\ref{pn1})-(\ref{pn3}) into (\ref{a11}) yields
\bea \lb{a13}
& &\sin^2 \frac{\phi_u}{2} = \frac{1}{2}  \frac{1 - \beta^{2}}{\left(1 + \bbe . \bN \right)\left(1 + \bbe . \bN' \right)}
\left\{ \left(1 - \bN . \bN'\right)\left[1 - 2(1 + \gamma)\frac{w}{c^2} - \bbe . \left(\bla + \bla'\right)\right] \right. \qquad \qquad \qquad   \nonumber \\
& &\left. \qquad \qquad \qquad \qquad \qquad \qquad \qquad \qquad \qquad \qquad + \sqrt{2\left(1 - \bN .\bN'\right)} \, \,  
\left( \bla' - \bla\right) . \, \bs  + O(c^{-4}) \right\} \, 
\eea

Hence, assuming $\phi_U^{(0)}\neq 0$ and taking the square root of the r.h.s. of Eq. (\ref{a13}) 
\be \lb{a14}
\sin \frac{\phi_{u}}{2} = \frac{\sqrt{1 - \beta^{2}}}{\sqrt{\left(1 + \bbe . \bN \right)\left(1 + \bbe . \bN' \right)}}
\left\{ \sin \frac{\phi_U^{(0)}}{2} \left[1 - (1 + \gamma)\frac{w}{c^2} - \frac{1}{2} \bbe . \left(\bla + \bla'\right)\right] + \frac{1}{2}   
\left(\bla' - \bla\right). \, \bs + O(c^{-4}) \right\} \, .
\ee

Using the results established in Sec. IV, it may be seen that the angle $\phi_{u}$ admits the expansion
\be \lb{a15}
\phi_{u} = \phi_{u}^{(0)} + \phi_{u}^{(PN)} + O(4)\, ,
\ee
where $\phi_{u}^{(0)}$ is defined by (\ref{an0}) and $\phi_{u}^{(PN)}$ is given by
\bea 
\phi_{u}^{(PN)}&=& \left[1 - \frac{\bbe . (\bN + \bN')}{ 1 + \bN . \bN'}\right] \left[\sqrt{\frac{2}{1 + \bN . \bN'}} \, (\bla' - \bla)\, . \, \bs   - 2(1 + \gamma) \frac{w}{c^2} \sqrt{\frac{1 - \bN . \bN'}{1 + \bN . \bN'}}\right] \nonumber \\
& &\mbox{}- \sqrt{\frac{1 - \bN . \bN'}{1 + \bN . \bN'}}\, \bbe \, . \, (\bla + \bla')  \, . \lb{a16}
\eea

This formula yields a thorough determination of the influence of the motion of the observer and of the gravitational light deflection on the angular distance described within the post-Newtonian approximation.

\section{Concluding remarks}

This work yields a general, rigorously covariant determination of the angular distance between two point sources of light as measured by an observer moving with a given velocity. The fundamental formulae (\ref{an8}) and (\ref{cov}) show that the theoretical calculation of the angular distance can be carried out when the so-called reception time transfer function is known throughout space-time. So it can be said that the problem of space astrometry involves one and only one unknown function. Furthermore, it must be pointed out that our formulae do not involve any tetrad formalism to define the directions of light rays. It is also worthy of note that the aberration and the gravitational deflection of light are not separately tackled. 
	
The case of weak gravitational fields is explicitly and thoroughly treated within the post-linear approximation. The formulae (\ref{phi})-(\ref{D2}) yield  the perturbative expansion of the angular separation up to the second order in $G$. These results will enable to discuss the effects which could be observed by the next generation of stringent tests of general relativity like LATOR or ASTROD. Finally, the expression of the angular separation is obtained within the post-Newtonian approximation truncated at the order $1/c^3$. We are currently using this last calculation for determining the contributions of the mass-multipole coefficients of the gravitational field to the bending of light in the vicinity of a giant planet like Jupiter.

\end{document}